\documentclass[galaxies,accept,moreauthors,galaxies,pdftex,10pt,a4paper]{mdpi}

\firstpage{1}
\articlenumber{57}
\doinum{10.3390/galaxies6020057}
\pubvolume{6}
\pubyear{2018}
\copyrightyear{2018}
\history{Received: 18 March 2018; Accepted: 20 May 2018; Published: 24 May 2018}
\pdfoutput=1

\usepackage{amssymb}

\newcommand{\mn}{{Mon. Not. R. Astron. Soc.}}
\newcommand{\mnras}{\mn}
\newcommand{\apj}{{Astron. J.}}
\newcommand{\apjl}{{Astrophys. J. Lett.}}

\newcommand{\aap}{{Astron. Astrophys.}}

\newcommand{\prd}{\mbox{Phys. Rev. D}}
\newcommand{\nat}{Nature}

 \theoremstyle{mdpi}
 \newcounter{thm}
 \newcounter{ex}
 \newcounter{re}
\Title{Probing Black Hole Magnetic Fields with QED}

\Author{Ilaria Caiazzo $^{*}$ and Jeremy Heyl}

\AuthorNames{Ilaria Caiazzo and Jeremy Heyl}

\address[1]{%
Department of Physics and Astronomy, 6224 Agricultural Road, University of British Columbia, \mbox{Vancouver, BC V6T 1Z1, Canada}}

\corres{\hangafter=1 \hangindent=1.05em \hspace{-0.82em}Correspondence: ilariacaiazzo@phas.ubc.ca; Tel.: +1-604-822-0995}

\abstract{The effect of vacuum birefringence is one of the first predictions of quantum electrodynamics (QED): the presence of a charged Dirac field makes the vacuum birefringent when threaded by magnetic fields. This effect, extremely weak for terrestrial magnetic fields, becomes important for highly magnetized astrophysical objects, such as accreting black holes. In the X-ray regime, the polarization of photons traveling in the magnetosphere of a black hole is not frozen at emission but is changed by the local magnetic field. We show that, for photons traveling along the plane of the disk, where the field is expected to be partially organized, this results in a depolarization of the X-ray radiation. Because the amount of depolarization depends on the strength of the magnetic field, this~effect can provide a way to probe the magnetic field in black-hole accretion disks and to study the role of magnetic fields in astrophysical accretion in general. }


\begin{document}




\section{Introduction}

In the theory of accretion disks around black holes and astrophysical accretion in general, magnetic fields play a crucial role. They are expected to be the main source of shear stresses, without which accretion cannot occur \cite{1973A&A....24..337S,1991ApJ...376..214B}. Moreover, magnetic fields in the inner regions of black-hole accretion disks are thought to lead to the formation of relativistic jets through the Penrose--Blandford--Znajek mechanism \cite{1977MNRAS.179..433B,2011MNRAS.418L..79T}. However, information on the strength and structure of magnetic fields around black holes is hard to obtain by direct observations. From the analysis of the spectra of two Galactic stellar-mass black holes, Miller {et al.}~\cite{2006Natur.441..953M,2008ApJ...680.1359M,2016ApJ...821L...9M}  showed that a wind is generated by magnetic processes as close as 850 GM/c$^2$ to the hole. They also obtained an estimate of the strength of the magnetic field when a certain magnetic process is assumed \cite{2016ApJ...821L...9M}. The only indication that we have on the magnetic field structure closer to the central engine comes from interferometry observations of the radio polarization from Sagittarius A*, the supermassive black hole at the center of the Milky Way, which shows evidence for a partially ordered magnetic field on scales of ~12 GM/c$^2$ \cite{2015Sci...350.1242J}. In this paper, we describe how X-ray polarization measurements from black-hole accretion disks could provide a way to probe, for the first~time, the strength and structure of the magnetic field close to the event horizon.

If only classical electrodynamics is considered, at energies higher than 1--2 keV, the polarization of a photon emitted by the accretion disk is not affected by the presence of a magnetic field. The~linear polarization of X-ray photons stays the same as they travel through the magnetosphere of the hole all the way to the observer. At lower photon energies, the presence of a magnetized corona could destroy the linear polarization of X-ray photons due to the effect of plasma birefringence \cite{1985ApJ...298..147M,2009ApJ...703..569D,2018PhRvD..97h3001C}. In~quantum electrodynamics (QED),  the vacuum is also expected to be birefringent in presence of a~magnetic field. This effect, which was one of the first predictions of QED, has never been proven. Recent observations of the visible polarization from a radio-quiet neutron star \cite{2017MNRAS.465..492M} have strongly hinted that vacuum birefringence is indeed affecting the photons' polarization. If the vacuum is indeed birefringent, after~photons are emitted from the disk, their polarization will change as they travel through the magnetized vacuum.

In classical electrodynamics, photons do not interact with other electromagnetic fields as Maxwell equations are linear in the fields. In QED, the presence of a Dirac current in the vacuum results in an~addition to the usual action integral of the electromagnetic field that is more than quadratic in the fields. This implies that the interaction between the fields is not linear as photons can interact with virtual electron--positron pairs as they travel through the magnetized vacuum. As a result, the~speed at which light travels through the vacuum depends on its polarization and on the strength of the field. \mbox{In other words}, in presence of a magnetic field the vacuum becomes birefringent, {i.e}., it acquires an index of refraction that is different depending on the angle between the direction of the photon's polarization and the magnetic field. A detailed derivation of the vacuum birefringence in QED is described by Heyl and Caiazzo, in this volume.

In this paper, we assume the strength of the magnetic field in the accretion disk to be the minimum needed for accretion to occur if an $\alpha$-model structure of the disk is considered. We find that the effect of vacuum birefringence on the photon polarization becomes important, depending on the angular momentum of the black hole and that of the photon, around 10 keV, for both stellar-mass and supermassive black holes. A stronger (weaker) field would shift this range to lower (higher) energies. Observation of the X-ray polarization from accretion disks in the 1--30 keV range, if properly modeled with QED, would both probe the strength of the magnetic field and test the currently accepted models of astrophysical accretion. Several observatories with an X-ray polarimeter on board are now at different stages of development: in the 1--10~keV range, the NASA SMEX mission \textit{IXPE} \cite{2016SPIE.9905E..17W} and the Chinese--European \textit{eXTP} \cite{2016SPIE.9905E..1QZ}; in the hard-X-ray range, 15--150 keV, the balloon-borne \textit{X-Calibur}~\cite{2014JAI.....340008B} and \textit{PoGO+} \cite{2018MNRAS.tmpL..30C} and Friis {et al.}, in this volume; and, in the sub-keV range, the narrow band (250 eV) \textit{LAMP}~\cite{2015SPIE.9601E..0IS} and the broad band (0.2--0.8 keV) rocket-based \textit{REDSox} \cite{SPIE_REDSoX}.

In Section \ref{sec:model}, we introduce our model and our assumptions and, in Section~\ref{sec:results}, we show the energy at which QED becomes important given our assumptions as a function of the black hole spin and we show the effect of vacuum birefringence on the polarization of X-ray photons traveling near the disk~plane, where we assume the magnetic field to be partially organized. For a more detailed derivation of our equations, please see \cite{2018PhRvD..97h3001C}.

\section{Model}
\label{sec:model}

Our calculations are performed in the Kerr metric surrounding a spinning black hole, with~spin parameter \emph{a} = J/(cM), ranging from $a=0$ (Schwarzschild black hole) to \emph{a} = GM/c$^2$ (critical spin). To calculate the strength of the magnetic field in the mid-plane of the disk, we have to model the structure of the inner disk. In particular, we have to make an assumption on what is generating the shear stresses needed for accretion to occur. We follow the $\alpha$-model, suggested first by \citet{1973A&A....24..337S}, for which tangential stresses between layers are generated by magnetic field and~turbulence:
\begin{equation}
  \label{eq:stress}
  t_{\hat{\phi}\hat{r}} = \rho c_s v_t + \frac{B^2}{4\pi} = \alpha P
\end{equation}
where $\rho$ is the mass density, $c_s$ is the speed of sound, $v_t$ is the turbulence velocity, $B$ is the magnetic field strength, $P$ is pressure and $t_{\hat{\phi}\hat{r}}$ is the shear stress as measured in a frame of reference moving with the gas. The last equality is called the $\alpha$-prescription, in which the efficiency of the angular momentum transfer is expressed with one parameter. Because the magnetic field is at the origin of the turbulence, we expect the two terms in Equation~(\ref{eq:stress}) to be of the same order, so we estimate the magnetic field strength to be $\sim (4\pi\alpha P)^{1/2}$.

To calculate the pressure in the mid-plane, we employ the disk structure equations in \citet{1973blho.conf..343N}, with the correction to the hydrostatic equilibrium obtained by \citet{1995ApJ...450..508R}. The general relativistic equations in the two papers are written as Newtonian values times relativistic corrections, the latter expressed by functions that are equal to one in the Newtonian limit. In this paper, we use the following relativistic corrections:
\begin{subequations}
\label{eq:GR}
\begin{align}
  \mathcal{A} =& 1 + a_\star^2/r_\star^2 + 2a_\star^2/r_\star^3 \\
  \mathcal{B} =& 1 + a_\star/r_\star^{3/2} \\
  \mathcal{C} =& 1 - 3/r_\star + 2a_\star/r_\star^{3/2} \\
  \mathcal{D} =& 1 -2/r_\star +a_\star^2/r_\star^2 \\
  \mathcal{N} =& 1 - 4 a_\star/r_\star^{3/2} + 3a_\star^2/r_\star^2
\end{align}
\end{subequations}
where $a_\star=ac^2$/GM, $M$ is the mass of the black hole, $r_\star=rc^2$/GM and $r$ is the distance from the black hole (more precisely, the circumferential radius). The first four come from \cite{1973blho.conf..343N} and the last~one,\mbox{ $\mathcal{N}$, corresponds to} the quantity called $C$ in \cite{1995ApJ...450..508R}.
We also assume the pressure to be dominated by radiation and the opacity to be dominated by electron scattering. This assumption applies to the inner region of the accretion disk, which is also where the magnetic field is stronger. Outer regions of the disk will have no influence on our calculations as the magnetic field there is weak. We find the square of the strength of the magnetic field in the mid-plane to be:
\begin{equation}
  \label{eq:B}
  B^2 = \frac{8\pi c}{3\kappa_{es}} \sqrt{\frac{GM}{r^3}} \frac{\mathcal{N}}{\mathcal{D}} \; .
\end{equation}
where $\kappa_{es}$ is the electron scattering opacity. For a 10 M$_\odot$ black hole, at the innermost stable circular orbit of the disk (ISCO, or $r_I$), this corresponds to
\begin{equation}
\label{eq:bfield}
	B^2 = (0.36 - 1.22 \times 10^8 \, \mathrm{G})^2 \left( \frac{M}{10 \, \mathrm{M}_\odot} \right)^{-1} \left(\frac{1+X}{2} \right)^{-1}
\end{equation}
where the first value is for $a_\star=0$ and the second is for $a_\star=0.999 $ (the value diverges for $a_\star=1$) and $X$ is the hydrogen mass fraction. This value is a crude estimate of the minimum magnetic field needed in the mid-plane for accretion to occur if an $\alpha$-model is assumed. Magnetohydrodynamics and shear box simulations show that the strength of the magnetic field decreases moving away from the mid-plane toward the photosphere. However, Equation~(\ref{eq:B}) reproduces both the strength and the scaling with distance of the magnetic field at the photosphere obtained by simulations \cite{2009ApJ...691...16H,2013ApJ...769..156S}, and of the minimum estimates obtained by \citet{2016ApJ...821L...9M}. We decided therefore to use the analytic expression in Equation~(\ref{eq:B}) as our best guess for the strength of the magnetic field at the photosphere.

To describe the evolution of the polarization of a single photon, we used the Poincar\'e formalism, in which the polarization is described by a unit vector $\mathbf{s} = (Q,U,V)/I$, \mbox{where $I,Q,U$ and $V$} are the Stokes parameters, and the polarization states for fully polarized light are mapped on the surface of a unit sphere. Following Kubo and Nagata~\cite{Kubo:81,Kubo:83}, the polarization of a wave in a birefringent medium evolves as:
\begin{equation}
  \label{eq:pol}
  \frac{\partial \mathbf{s}}{\partial x_3} = \hat{\mathbf{\Omega}} \times\mathbf{s} + (\hat{\mathbf{T}}\times\mathbf{s}) \times \mathbf{s}
\end{equation}
where $\hat{\mathbf{\Omega}}$ is the birefringent vector, $\hat{\mathbf{T}}$ is the dichroic vector and $x_3$ is the length of the photon path. In~the case of the QED vacuum with an external magnetic field to one-loop order and a weak electric field, $\hat{\mathbf{T}} = 0$ (there is no real pair production) and the amplitude of the birefringent vector $\hat{\mathbf{\Omega}}$ is proportional to the difference between the indices of refraction for the two polarization states: the one parallel to the magnetic field ($n_\parallel$) and the one perpendicular ($n_\perp$). Equations~(53) and~(54) of Heyl and Caiazzo, this~volume, yield for $B \ll B_\mathrm{QED}$:
\begin{equation}
  \label{eq:birivector}
  \hat{\Omega} = k_0(n_{\parallel} - n_{\perp}) = k_0\frac{\alpha_{\mbox{\tiny{QED}}}}{30 \pi} \left(\frac{B}{B_{\mbox{\tiny{QED}}}}\right)^2\sin^2\theta
\end{equation}
where $k_0 = 2 \pi \nu / c$ is the unperturbed wavenumber of the photon, $\theta$ is the angle between the direction of the motion of the photon and the external field,
$\alpha_{\mbox{\tiny{QED}}}$ is the fine structure constant and $B_{\mbox{\tiny{QED}}} = m_e^2 c^3/ (\hbar e) \simeq 4.4 \times 10^{13} $ G.

From Equations~(\ref{eq:B}) and~(\ref{eq:birivector}), we can find the magnitude of the birefringent vector as function of the distance from the hole along the plane of the disk. After assuming a structure for the magnetic field, we can integrate Equation~(\ref{eq:pol}) to find how the polarizations of photons traveling in the magnetosphere close to the disk plane evolve.

\section{Results}\vspace{-6pt}
\label{sec:results}
\subsection{Polarization-Limiting Radius}
Before calculating the evolution of the photon polarization for a defined structure of the magnetic field, it is interesting to look at the quantity called the polarization-limiting radius (PLR). The PLR provides an estimate for the distance from the black hole at which the vacuum birefringence stops affecting the photon polarization because the magnetic field has become too weak. From Equations~(\ref{eq:B}) and~(\ref{eq:birivector}) and Equation~(56) of Heyl and Caiazzo, this volume, we find the PLR for a black hole to be:
\begin{equation}
\label{eq:PLR}
\frac{r_p c^2}{GM} = \left(\frac{2k_0\hbar m_p}{15\pi  m_e^2 c (1+X)} \frac{ \mathcal{N}(r_p)}{\mathcal{D}(r_p)} \right)^2
\end{equation}
where $m_p$ is the mass of the proton and $m_e$ is the mass of the electron. From Equation~(\ref{eq:PLR}), we can derive the energy at which the PLR is equal to the ISCO.

\begin{figure}[H]
\centering
\includegraphics[width=0.6\textwidth]{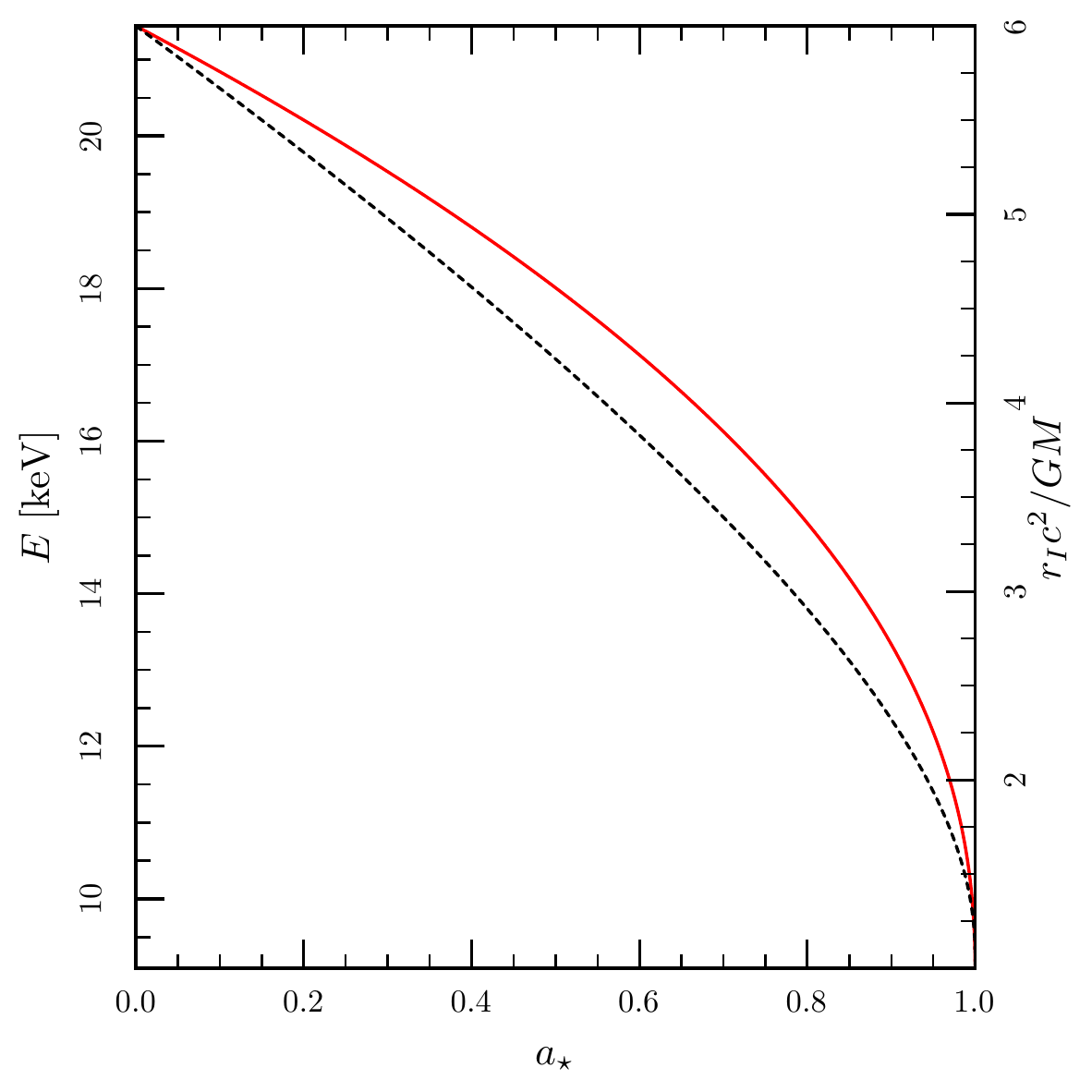}
\caption{The plot shows, on the left $y-$axis, the energy at which $r_p = r_I$ (\textbf{solid red line}).  On~the right $y-$axis, the ISCO for a black hole as function of the spin parameter $a_\star$ (\textbf{dashed black line}). \mbox{Figure from \cite{2018PhRvD..97h3001C}.}}
\label{fig:pol_rad}
\end{figure}

Figure~\ref{fig:pol_rad} shows the ISCO as a function of the black hole spin (black dashed line, right y-axis) and the photon energy at which the PLR is equal to the ISCO (solid red line, left y-axis).
Figure~\ref{fig:pol_rad} provides a rough estimate of the photon energy at which QED becomes important: if our estimate of the magnetic field strength is correct, for rapidly spinning black holes, the effect of QED will be important around a photon energy of 10 keV or lower, while for slowly spinning black holes, QED will affect the polarization only above 10--20~keV. However, if the magnetic field is stronger (weaker) the energy threshold will be lower (higher). This result does not depend on the mass of the black hole, so it holds for both stellar-mass and supermassive black holes. The PLR estimate does not take into account light bending: if a photon is emitted with large retrograde angular momentum, its path through the magnetosphere will be longer, so retrograde photons at lower energies can also be affected, as we find in the next section.

\subsection{Edge-on Photons}
To better understand how vacuum birefringence affects the polarization of photons traveling through the black hole magnetosphere, we assume a simple structure for the magnetic field threading the accretion disk, and we study how the polarization changes for photons traveling parallel to the disk plane. Recent observations of the radio polarization coming from the region close to the event horizon of Sagittarius A* suggest the presence of a partially organized field \cite{2015Sci...350.1242J}. It is reasonable to assume the magnetic field to be organized on some length-scale that reflects the competition between the magnetic field itself, which would tend to be organized, and the shear of the disk, which prevents big structures from forming. We therefore assume the disk to be divided into regions of constant magnetic-field direction, which is also the structure often assumed for the magnetic field in the plane of the disk by magnetohydrodynamics simulations \cite{2015MNRAS.446L..61P}.
We pick two different length-scales to test how our assumption on the size of the magnetic loops affects our results. Since we expect the length scale to be related to both the distance to the hole and to the size of the hole itself, we first divide the disk into five regions, each twice as large as the previous one: from the ISCO to twice the ISCO, \mbox{to 4 times} the ISCO, to 8 times the ISCO, to 16 times the ISCO, and to infinity. For simplicity, we call this configuration the \textit{2-fold configuration}. In the second configuration, the regions of constant magnetic-field \mbox{direction are each 1.5 times} as large as the previous one: from the ISCO to 1.5 times the ISCO, \mbox{to 2.3 times} the ISCO, to 5.1 times the ISCO, to 7.6 times the ISCO, to 11 times the ISCO, \mbox{to 17 times} the ISCO, and to infinity. For~simplicity, we call this configuration the \textit{1.5-fold configuration}.

We analyze the evolution of the polarization of single photons as they travel along geodesics through the magnetosphere. On the Poincar\'e sphere, their polarization will perform a random walk, where in each region the direction of the step is given by Equation~(\ref{eq:pol}) and the rotation angle around $\mathbf{\hat{\Omega}}$ is given by
\begin{equation}
\label{eq:maxmin}
\Delta\Theta =  E K \int \sin^2\theta r_\star^{-\frac{3}{2}}\frac{\mathcal{N}}{\mathcal{D}\mathcal{C}} \frac{(l_\star/r_\star^{3/2} - \mathcal{B})^2}{(l_\star^2(-1+2/r_\star)/r_\star^2-4l_\star a_\star/r_\star^3 + \mathcal{A})^{1/2}} dr_\star
\end{equation}
where $E$ is the energy of the photon at infinity, $l_\star$ is the dimensionless specific angular momentum of the photon ($l_\star = L/E\times c^2$/(GM)) and  $K = m_p/[15\pi m_e^2c^2(1+X)]$. Since we are \mbox{considering photons traveling close} to the equatorial plane, general relativity does not affect their polarization's direction.

\begin{figure}[H]
\begin{minipage}{0.5\linewidth}
		\centerline{\includegraphics[width=\textwidth]{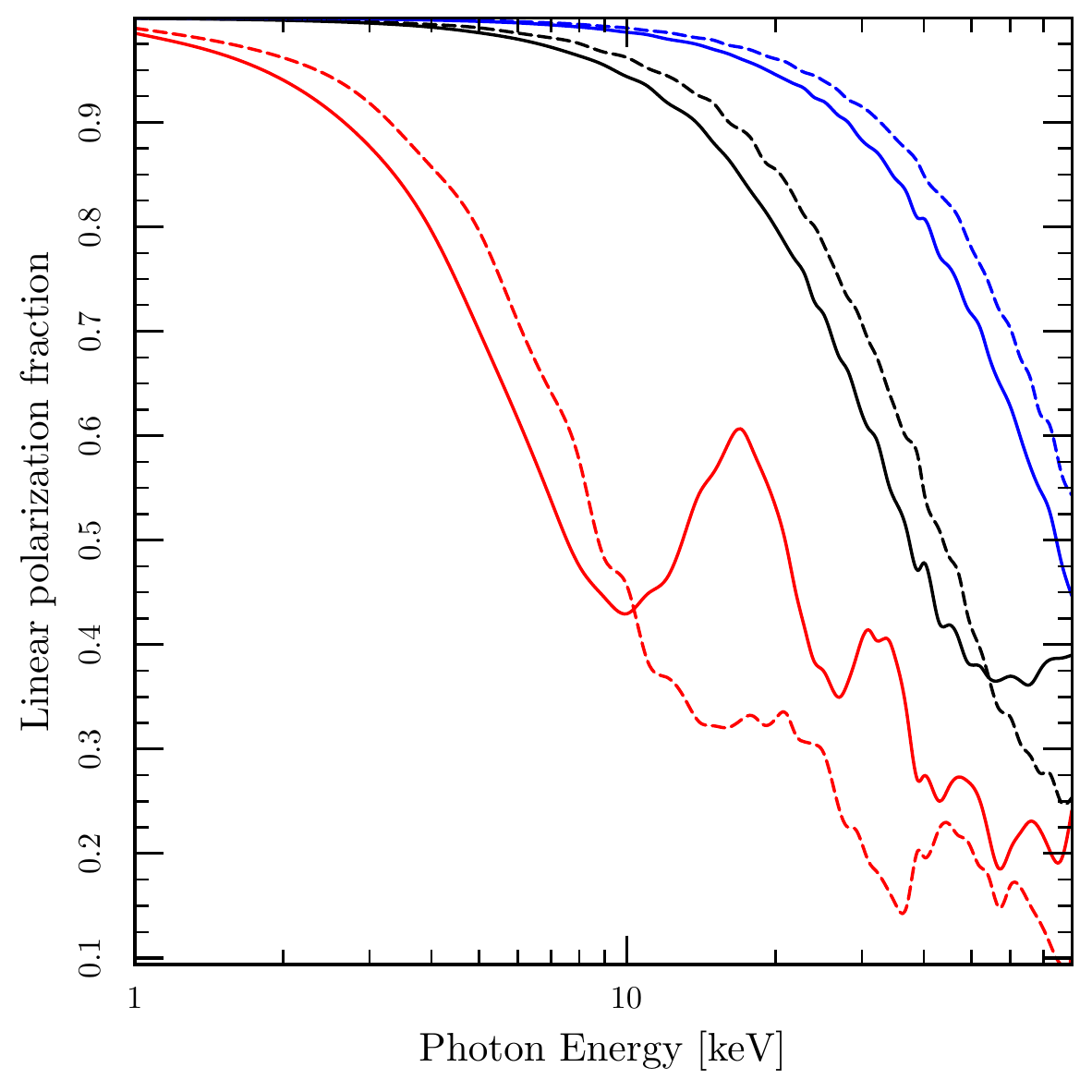}}
		\centerline{(\textbf{a})}
	\end{minipage}
	\hfill
	\begin{minipage}{0.5\linewidth}
		\centerline{\includegraphics[width=\textwidth]{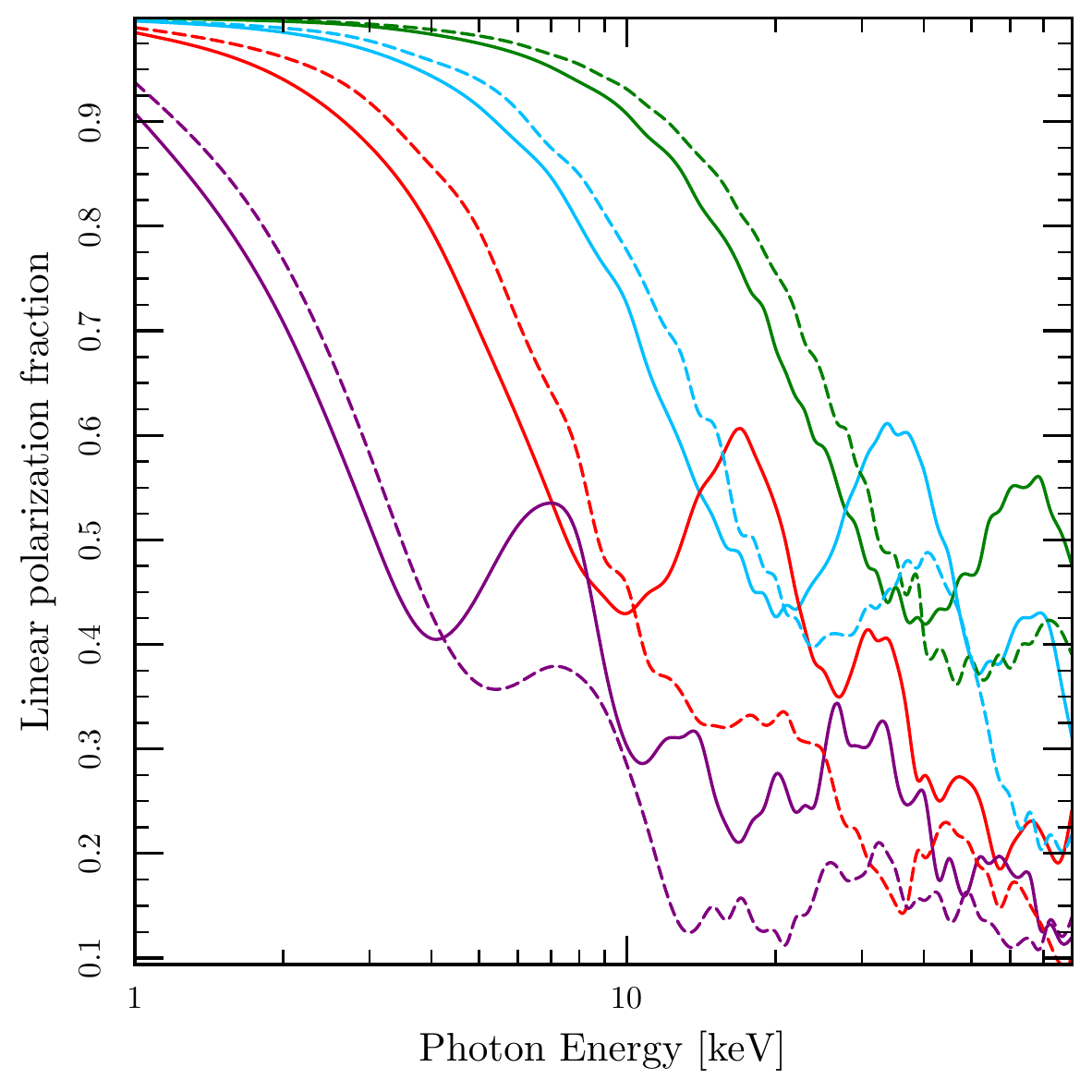}}
		\centerline{(\textbf{b})}
	\end{minipage}
	\vfill
\caption{Final polarization fraction vs.\ photon energy calculated in the \textit{2-fold configuration} (solid lines) and in the \textit{1.5-fold configuration} (dashed lines): (\textbf{a}) left to right, maximum retrograde (90\% $l_-$) angular momentum photons (red), zero angular momentum photons (black) and maximum prograde (90\% $l_+$) angular momentum photons (blue), coming from the ISCO of a black hole with $a_\star=0.9$; and (\textbf{b})~ 90\% $l_-$ photons for, left to right, $a_\star=0.99$ (purple), 0.9 (red), 0.7 (light blue) and 0.5~(green). Figure from \cite{2018PhRvD..97h3001C}.}
\label{fig:plots}
\end{figure}

We perform a Monte-Carlo simulation for 6000 photons, calculating the evolution of their polarization from the ISCO to infinity. Each of the 6000 photons is emitted with the same specific angular momentum $l$ and the same energy at infinity $E$ from the ISCO of a black hole with spin parameter $a_\star$. We take the angle between the magnetic field and the photon, $\theta$, and the angle between \mbox{\textbf{s} and $\hat{\mathbf{\Omega}}$} to be constant in every region, and we take their values as random in each run. We then take the average of the linear polarization over the 6000 photons. We repeat the same calculation for photons with different specific angular momenta: zero angular momentum photons ($l=0$), photons initially rotating with the disk at 90\% the maximum prograde angular momentum ($l=0.9\, l_+$) and photons initially rotating against the disk at 90\% the maximum retrograde angular momentum ($l=0.9\, l_-$). We also employ different photon energies between 1 and 80 keV and for four different spins of the hole: $a_\star = 0.5$, 0.7, 0.9 and 0.99.  The results are shown in Figure~\ref{fig:plots}.

In Figure~\ref{fig:plots}, the dashed lines show the results for the \textit{1.5-fold configuration} and the solid lines show the results for the \textit{2-fold configuration}. We find that, if magnetic loops are smaller, the depolarization effect is reduced linearly with the size of the loops: in our example, the dashed lines fall on top of the solid lines if we rescale them by 2/1.5. However, the solid lines show peaks that are not present in the dashed lines. For example, for a hole rotating with spin $a_\star = 0.99$  in the \textit{2-fold configuration} (purple solid line, right panel) the polarization fraction peaks at 7 keV and then again at 14 keV, at 21 keV and so on. These peaks are due to the fact that at those energies the integral in Equation~(\ref{eq:maxmin}) reaches, in the first zone of the disk, an average value of $\pi$, and therefore, the polarization vector remains closer to the $S_1-S_2$ plane. In the \textit{1.5-fold configuration}, this does not happen because the first region is smaller and the second region has a bigger effect on the final polarization, washing out the peaks. Ideally, the~presence of features in the polarization spectrum such as the peaks shown for the \textit{2-fold configuration} could provide hints on the structure of the magnetic field in the disk.

All of the aforementioned results are independent of the black hole mass.

\subsection{A Simulation for GRS 1915+105}
\label{sec:grs}
To understand whether the upcoming polarimeters will be sensitive to the effects of QED, we simulated the observed polarization of the black-hole binary GRS 1915+105 with \textit{eXTP} \mbox{and \textit{IXPE}}. GRS 1915+105 is a bright microquasar that hosts a rapidly spinning black hole.  Measurements of its spin, which rely on observations in both X-rays and optical, seem to indicate a spin parameter $a_\star \gtrsim 0.98$ \cite{2006ApJ...652..518M,2013ApJ...775L..45M}. We assume an inclination angle of $75^\circ$ \cite{1994Natur.371...46M,1999MNRAS.304..865F}, and we use the polarization spectra from Figure~7 of Schnittman~and~Krolik (2009) \cite{2009ApJ...701.1175S}.  To calculate the effects of the vacuum birefringence, we assume that the bulk of the radiation comes from near the ISCO and has zero angular momentum.  To simulate the response of the instruments, we employ the code XIMPOL \cite{2016cosp...41E.129B}.

\begin{figure}[H]
\begin{minipage}{0.5\linewidth}
		\centerline{\includegraphics[width=\textwidth]{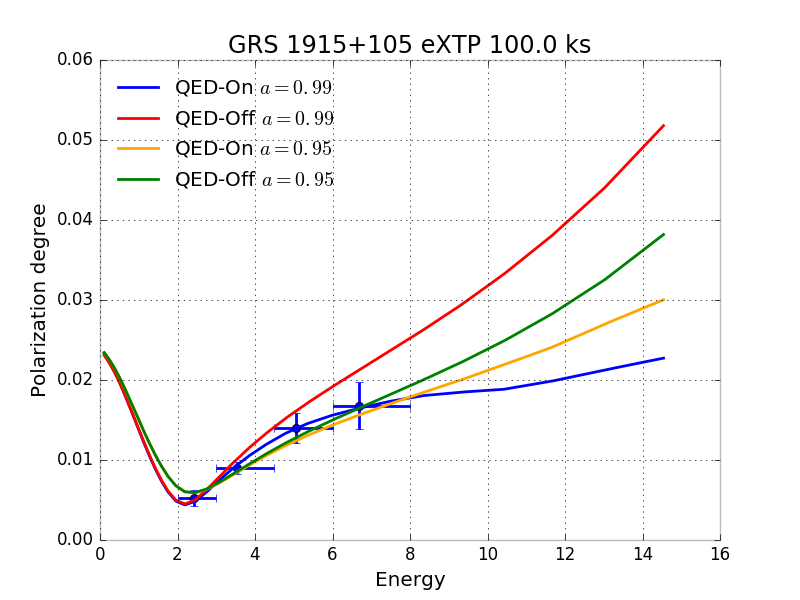}}
		\centerline{(\textbf{a})}
	\end{minipage}
	\hfill
	\begin{minipage}{0.5\linewidth}
		\centerline{\includegraphics[width=\textwidth]{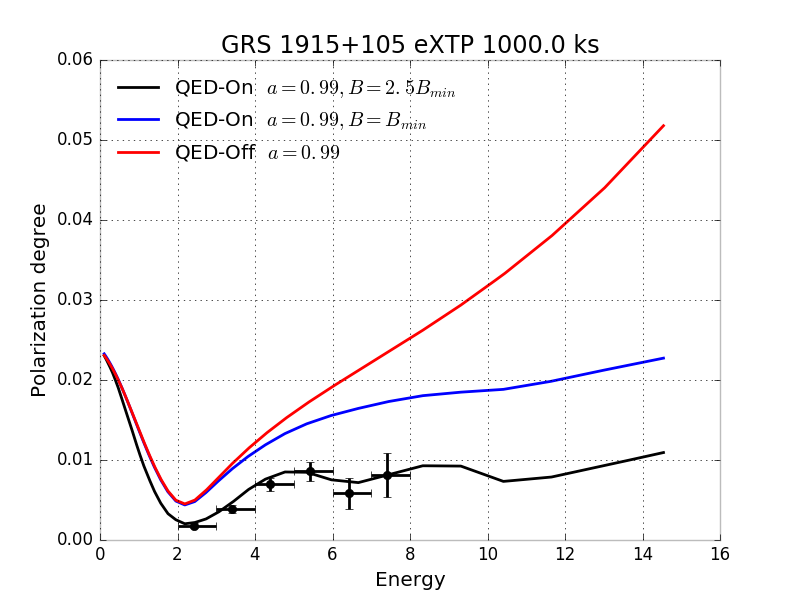}}
		\centerline{(\textbf{b})}
	\end{minipage}
	\vfill
\caption{Observed polarization degree for the black-hole binary GRS 1915+105. (\textbf{a}) Model~with $a_\star=0.99$ with QED (blue line) and without QED (red line); model with $a_\star=0.95$ with QED (yellow~line) and without QED (green~line). Blue dots are a simulated 100~ks observation with \textit{eXTP} \mbox{for the blue line model}.  (\textbf{b}) Model with $a_\star=0.99$ with QED and the minimum magnetic field (blue~line) and without QED (red line); model with $a_\star=0.99$ with QED and 2.5 times the \mbox{minimum magnetic field} (black~line). Black dots are a simulated 1~Ms observation with \textit{eXTP} for the black line~model.}
\label{fig:ximpol}
\end{figure}

Figure~\ref{fig:ximpol}a shows the observed polarization degree for two spin parameters, $a_\star=0.95$ and $a_\star=0.99$, both with and without including QED. The blue dots show a simulated 100~ks observation with \textit{eXTP} (which would correspond to approximately 300~ks with \textit{IXPE}), assuming the emission model to be the one with $a_\star=0.99$ and with QED (blue line). We can immediately see that QED has an effect in the energy range of the upcoming polarimeters (2--8 keV). In addition, if QED were not included in the model, it would be easy to mistake a black hole actually spinning at $a_\star=0.99$ (blue line) with one spinning at $a_\star=0.95$ (green line). In the left panel of Figure~\ref{fig:ximpol}, all the models are calculated assuming the minimum magnetic field needed for accretion to occur in an $\alpha-$model (Equation (\ref{eq:B})). In Figure~\ref{fig:ximpol}b, we show the effect of a stronger magnetic field. The red and the blue line are the same as in Figure~\ref{fig:ximpol}a: $a_\star=0.99$ and the minimum magnetic field, with and without~QED, while~the black line represents a model with the same parameters but a magnetic field 2.5 times stronger. The~black dots show a simulated 1~Ms observation with \textit{eXTP} ($\sim$3 Ms with \textit{IXPE}). We can see that the curves are very different, with the QED effect being much stronger for the stronger magnetic field, and that the peaks have shifted into the 2--8~keV range. Of course, the magnetic field structure that we use in this paper is just a toy model, but the peaks show that the QED effect can be sensitive to the magnetic field structure, and the upcoming polarimeters would be sensitive enough to detect~them.

We want to stress that these figures show preliminary calculations, and further work is required to model the expected polarization degree with \textit{IXPE} and \textit{eXTP}. Indeed, our model assumes the flux to be dominated by photons coming from close to the ISCO and with nearly zero angular momentum, which could be a good assumption for high-energy photons but the contribution of photons coming from more distant regions has to be properly included in the calculations for low-energy photons. Moreover, the structure of the magnetic field that we employ is just a simple toy model, and better calculations are needed to make a prediction on whether features like the peaks in the polarization degree would be detectable and at which energies they would be present.

\section{Discussion}

In Figure~\ref{fig:plots}, all photons were emitted with the same polarization. If the vacuum were not birefringent, their final polarization would still be the same, and the final linear polarization fraction would still average at one. We can therefore conclude that vacuum birefringence has a big impact on the polarization of X-ray photons, especially for fast-spinning black holes and for red-shifted (retrograde) photons. The reason the effect is stronger for higher spinning parameters is because the ISCO is closer to the event horizon and, therefore, the magnetic field is stronger, but also because photons perform more orbits around fast-spinning holes, staying longer in the strong magnetic field~region. Retrograde photons are more affected for two reasons: they perform more orbits around the black hole with respect to zero angular momentum and prograde photons, and they receive a red-shift, which~means that their energy at emission was higher.

The results shown in Figure~\ref{fig:plots} were obtained for the minimum magnetic field needed to generate enough shear stresses for accretion to occur in an $\alpha$-model for the accretion disk. The actual magnetic field threading the accretion disk could be higher, leading to a stronger effect of the vacuum birefringence on the polarization. In general, a stronger (or weaker) magnetic field would shift the $x-$axis of Figure~\ref{fig:plots} to a lower (higher) energy range, and the shifting would scale with the square of the magnetic field, as shown in Figure~\ref{fig:ximpol}.

The simulations presented in Section~\ref{sec:grs} are not intended to be predictive as more detailed models are required for the structure of the magnetic field close to the disk plane and for the contribution to the total emission from photons emitted at different distances to the central engine. However, they~show that vacuum birefringence has an effect on the observed polarization of fast-spinning black holes that can be detected in the energy range of the upcoming polarimeters \textit{IXPE} and \textit{eXTP}.

Our analysis is restricted to edge-on photons, traveling close to the disk plane, where we expect the magnetic field to be partially organized on small scales. Further studies are needed to calculate the effect of vacuum birefringence for photons coming out of the disk plane, where we expect the magnetic field to be organized on large scales. In this case, the effect of QED could be the opposite of what happens for edge-on photons: the organized magnetic field could align the polarization of photons traveling through the magnetosphere, resulting in a larger net observed polarization.

\section{Materials and Methods}

The Monte-Carlo simulations were performed by numerically integrating Equation~(\ref{eq:maxmin}) in Maple. A detailed derivation of the equations can be found in \cite{2018PhRvD..97h3001C}. The simulations for \textit{eXTP} in Section~\ref{sec:grs} were performed using the code XIMPOL \cite{2016cosp...41E.129B}.


\vspace{6pt}


\authorcontributions{I.C.\ and J.H.\ conceived the calculations for this paper.  I.C.\ performed the calculations under the supervision of J.H.\ and wrote the bulk of the paper.}

\acknowledgments{We thank the anonymous referees for useful suggestions that improved the paper significantly. We used the NASA ADS service, arXiv.org and SIMBAD. This work was supported by a Discovery Grant from the Natural Sciences and Engineering Research Council of Canada, the Canadian Foundation for Innovation and the British Columbia Knowledge Development Fund. I.C.\ is supported by a Four-Year-Fellowship from the University of British Columbia.}


\conflictofinterests{The authors declare no conflict of interest.}

\abbreviations{The following abbreviations are used in this manuscript:\\

\noindent
\begin{tabular}{@{}ll}
QED& quantum electrodynamics\\
ISCO& innermost stable circular orbit\\
PLR& polarization-limiting radius \\
IXPE& Imaging X-ray Polarimetry Explorer \\
eXTP& enhanced X-ray Timing and Polarimetry Mission \\
LAMP& Lightweight Asymmetry and Magnetism Probe\\
REDSox& The sounding Rocket Experiment Demonstration of a Soft X-ray Polarimeter \\
\end{tabular}}



%

\reftitle{References}


\end{document}